# Results from a Non-Perturbative Renormalization of Lattice Operators

G. Martinelli[a][b], C. Pittori[c] *, C.T. Sachrajda[d], M. Testa[a] and A. Vladikas[e]

[a]Dipartimento di Fisica, Università di Roma 'La Sapienza' and INFN, Sezione di Roma I,
P.le A. Moro, I-00185 Rome, Italy.

[b]Theory Division, CERN, CH-1211 Geneva 23, Switzerland.

[c]L.P.T.H.E., Université de Paris Sud, Centre d'Orsay, 91405 Orsay, France.

[d]Department of Physics, The University, Southampton SO9 5NH, U.K.

[e]Dipartimento di Fisica, Università di Roma 'Tor Vergata' and INFN, Sezione di Roma II,
Via della Ricerca Scientifica 1, I-00133 Rome, Italy.

We propose a general renormalization method, which avoids completely the use of lattice perturbation theory. We present the results from its numerical applications to two-fermion operators on a $16^3 \times 32$ lattice, at $\beta = 6.0$.

## 1. Introduction

Following the ideas presented last year at this conference [1], we have pursued a theoretical and numerical study, described in detail in [2], on the feasibility of a non-perturbative renormalization of generic composite operators in lattice QCD. We propose a renormalization method which avoids completely the use of lattice perturbation theory, known to be an important source of uncertainty in the extraction of physical results. This method, applicable to any general composite operator, is particularly useful when other non-perturbative techniques, based on Ward identities [3] or on renormalization conditions on hadron states [4], are not viable. A similar, but limited attempt, can be found in [5]. Here we present the main results of numerical application to two-fermion operators. Further applications to the four-fermion operators of the effective weak hamiltonian, and to heavy-light currents in the heavy quark effective theory, are under way [6].

## 2. Non-Perturbative Renormalization

Our renormalization method consists in mimicking on the lattice what is usually done in perturbation theory. One may fix the renormalization conditions of a certain operator by imposing that suitable Green functions, computed between off-shell quark and gluon states, in a fixed gauge, at a scale $\mu$, coincide with their tree level value. For illustrative purposes, we consider specific applications to matrix elements of two-quark operators. We define the renormalized operator $O(\mu) = Z_O O(a)$, by introducing the renormalization constant $Z_O$, which is found, by imposing the non-perturbative renormalization condition

$$Z_O\left(\mu a, g(a)\right) Z_\psi^{-1}\left(\mu a, g(a)\right) \Gamma_O(pa)|_{p^2=\mu^2} = 1. \quad (1)$$

$\Gamma_O$ is the forward amputated Green function[2] of the bare operator, computed between off-shell quark states with four-momentum $p$, with $p^2 = \mu^2$, in the Landau gauge. $Z_\psi$ is the quark wave function renormalization, which can be non-perturbatively defined from the conserved vector current, or from the quark propagator ($Z'_\psi$). This scheme solves the problem of large corrections in lattice perturbation theory, which are automatically included in the renormalization constants. The renormalized operator is independent of the regularization scheme; it depends, however, on the external states and on the chosen gauge. Matching with the corresponding operators, in some other standard continuum renormalization scheme (e.g. $\overline{MS}$), then requires a *continuum* perturbative calculation only, which is a common step in all approaches, standard or "boosted" perturbation theory [7] and non-perturbative renormalization. Our proposal is expected to work, whenever it is possible to fix the virtuality of the external states $\mu$ so that $\Lambda_{QCD} \ll \mu \ll 1/a$, in order to keep under control both non-perturbative and discretization effects. If such a "window" does not exist at current $\beta$ values, an accurate matching becomes impossible, not only in our approach, but also with any other method.

---

*Talk presented by C. Pittori.

[2]Suitably projected on the tree-level, and traced over colour and spin [2].

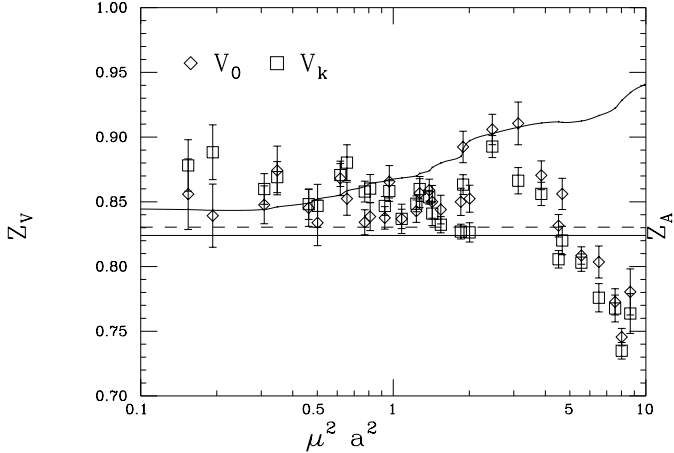

Figure 1. $Z_V$ as a function of $\mu^2 a^2$.

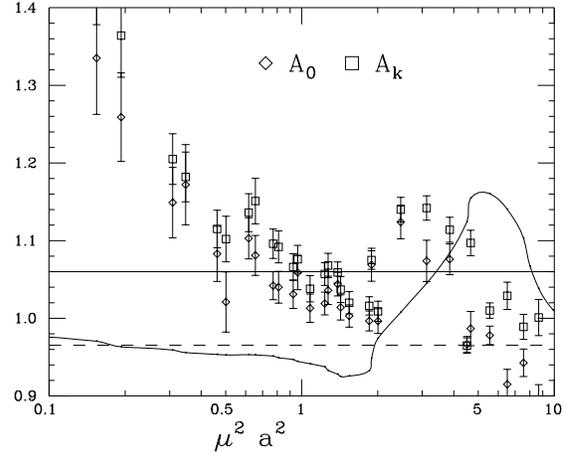

Figure 2. $Z_A$ as a function of $\mu^2 a^2$,

### 3. Numerical Results

In order to study the $\mu$-dependence of the renormalization constants, we have performed a simulation with 36 quenched configurations, on a $16^3 \times 32$ lattice, at $\beta = 6.0$, in the Landau gauge. To reduce discretization errors, we consider improved operators with the Clover fermion action, at $K = 0.1425$ ($am_q \sim 0.07$). Our non-perturbative (NP) results are shown in figs.1-4. We compare them with "Boosted Perturbation Theory" (BPT), using an effective coupling $\alpha_s^V \simeq 1.68\, \alpha_s^{LATT}$. To monitor the distortions due to discretization, the perturbative computation has also been performed on a lattice of the same size as the non-perturbative one, with a fixed spacing. We call this procedure "Boosted Discrete Perturbation Theory" (BDPT). In the figures the dashed line is from BPT, the curve is from BDPT, and the straight line is from the Ward identities (WI) method.

$\underline{Z_V}$: as expected, it is independent from the scale, within statistical errors, up to large values of $\mu^2$, where discretization errors become important. By using the points at $\mu^2 a^2 \sim 1$, corresponding to $\mu \sim 2$ GeV, we get $Z_V^{NP} = 0.84(1)$, to be compared with $Z_V^{WI} = 0.824(2)$ and $Z_V^{BPT} = 0.83$. The three methods are in good agreement.

$\underline{Z_A}$: it is also finite at all orders, and can be determined from WI. However, in the axial case (as in the pseudoscalar case) we expect to find a non-perturbative contribution from the pion pole, vanishing for large $\mu^2$ [2]. We interpret the strong $\mu$-dependence of $Z_A$, at low $\mu^2$, as the non-perturbative effect of the pseudoscalar state. There is no clear plateau between the non-perturbative regime and the large $\mu$ region. Nevertheless, it flattens around $\mu^2 a^2 \sim 1$, just before lattice artefacts become large in DBPT, and we get $Z_A^{NP} = 1.04(3)$, which agrees with the value $Z_A^{WI} = 1.06(2)$. BPT gives $Z_A$ smaller than one, indicating a $\sim 15\%$ higher orders effect.

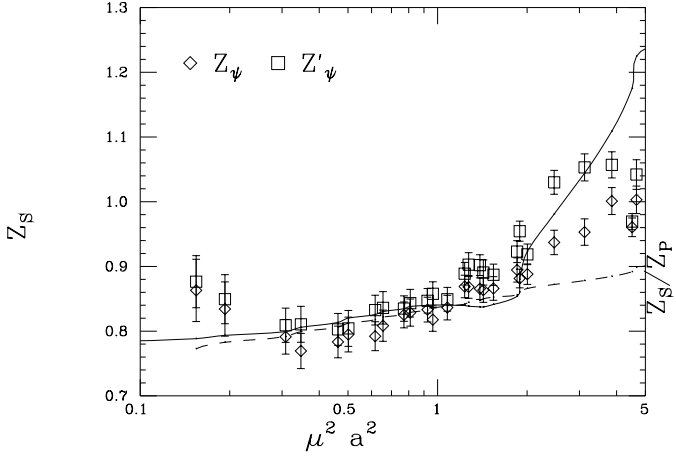

Figure 3. $Z_S$ as a function of $\mu^2 a^2$.

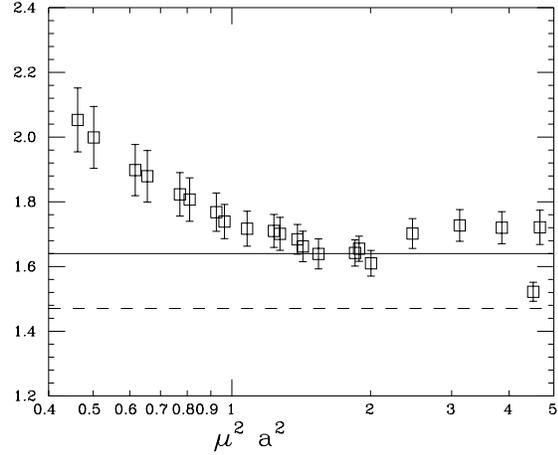

Figure 4. $Z_S/Z_P$ as a function of $\mu^2 a^2$.

$\underline{Z_S}$: this is an ideal quantity to check the validity of our method, since it cannot be determined using WI, and it is not affected by the pion pole. It is logarithmically dependent on the scale, and gauge dependent. We do not expect agreement with BPT, either at low $\mu^2$, due to higher order effects, or at high $\mu^2$, where lattice distortions are important, as shown by BDPT. NP results follow the theoretical expectations and we find good agreement with BPT in the range $0.3 \leq \mu^2 a^2 \leq 1$.

$\underline{Z_P}$: the pseudoscalar density (not shown here) is coupled to the pion, at low $\mu^2$; moreover it has large one-loop perturbative corrections, of order $\sim 35\%$. It is then not surprising that the NP result lies well below the BPT value, because of higher orders, not accounted for by one-loop BPT. With standard perturbation theory, the discrepancy would have been even worse.

$\underline{Z_S/Z_P}$: this ratio has a behaviour similar to $Z_A$. At $\mu^2 a^2 \sim 1$, the value obtained is in agreement with that determined by WI, whereas, given the result for $Z_S$, the significant difference with BPT should be due to $Z_P$. In conclusion, our numerical study suggests that there does exist a window, $1.1 \leq \mu \leq 2$ GeV, where the method can be applied, already at $\beta = 6.0$, and we expect the useful range to get larger, as $\beta$ increases.

We thank S. Petrarca for an early partecipation to this work.

## REFERENCES

1. G. Martinelli et al., Nucl. Phys. B (Proc. Suppl.) 34 (1994) 507.
2. G. Martinelli et al., CERN-TH.7342/94, submitted to Nucl. Phys. B.
3. L. Maiani and G. Martinelli, Phys. Lett. B178 (1986) 265; G. Martinelli et al., Phys. Lett. B311 (1993) 241.
4. M.B. Gavela et al., Phys. Lett. B211 (1988) 139.
5. C. Bernard et al., Nucl. Phys. B (Proc. Suppl.) 17 (1990) 593; 20 (1991) 410.
6. G. Martinelli et al., work in progress.
7. G.P. Lepage and P.B. Mackenzie, Phys. Rev. D48 (1993) 2250.